\documentclass[useAMS,usenatbib]{mn2e_lett}

\usepackage{graphicx}

\def \dn {\Delta_n}
\def \nupl {\nu_{\rm pl}}
\def \nucr {\nu_{\rm crv}}
\def \bb {\beta_b}
\def \imax {I_{\rm max}^{\rm bfc}}
\def \thmax {\theta_{\rm max}}
\def \psimax {\psi_{\rm max}}
\def \dbfc {\Delta_{\rm bfc}}
\def \iperp {I_{{\rm crv,}\perp}}
\def \omode {$\parallel$-mode}
\def \emode {$\perp$-mode}
\def \dmdn {\delta_{\rm mdn}}
\def \bfb {\delta_{\rm cut}}
\def \chisq {\chi^2/{\rm dof}}

\def \mref#1{(\ref{#1})}

\def \jtt {{\rm J}1012$+$5307\ }
\def \jttns {{\rm J}1012$+$5307}

\newcommand{\rlc}{R_{\rm lc}}

\def \nuobs {\nu_{\rm obs}}

\title[The nature of pulsar radio emission]{The nature of pulsar radio emission}

\author[J.~Dyks, B.~Rudak, 
and P.~Demorest
]{J. Dyks$^{1}$,
B.~Rudak$^{1}$, 
 and P. Demorest$^2$\\
$^{1}$Nicolaus Copernicus Astronomical Center, Toru\'n, Poland\\
$^{2}$Department of Astronomy, University of California, Berkeley, CA
94720-3411
}
\begin{document}

\date{Accepted .... Received 2009 August 10; in original form 2009 July 14}


\maketitle

\label{firstpage}

\begin{abstract}
High-quality averaged radio profiles of some pulsars exhibit
double, highly symmetric features both in emission and absorption.
It is shown that both types of features
are produced by a split-fan beam of 
extraordinary-mode curvature radiation (CR) that is emitted/absorbed by
radially-extended streams of magnetospheric plasma.
With no emissivity in the plane of the stream, such a beam produces
bifurcated emission components (BFCs) when our line of sight passes through the
plane.
A distinct example of double component created in that way is present
in averaged profile of a 5 ms pulsar \jttns.
We show that the component can indeed be very well fitted by the textbook 
formula
for the non-coherent beam of curvature radiation in the polarisation state
that is orthogonal to the plane of electron trajectory.
The observed  width of the BFC decreases with increasing frequency
at the rate that confirms the curvature origin.
Likewise, the double absorption
features (double notches) are produced by the same beam of the
extraordinary-mode CR, when it is eclipsed by thin plasma streams.
The intrinsic property of CR to create bifurcated fan beams
explains the double features in terms of very natural geometry
and implies the curvature origin of pulsar radio emission.
Similarly, ``double conal" profiles of class D
are due to a cut through a wider stream with finite extent in magnetic 
azimuth. Therefore, their width reacts very slowly to changes 
of viewing geometry due to the geodetic precession.
The stream-cut interpretation implies 
highly nonorthodox origin of both the famous S-swing of polarisation angle,
and the low-frequency pulse broadening in D profiles.
Azimuthal structure of polarisation modes in the CR beam 
allows us to understand
the polarised `multiple imaging' and
the edge depolarisation of pulsar profiles.
\end{abstract}


\section{Introduction}

Double `absorption' features in radio-pulse profiles 
were first identified in radio data independently
by Rankin \& Rathnasree (1997; B1929$+$10) and 
Navarro, Manchester, Sandhu, et al.~(1997; J0437$-$4715). 
McLaughlin \& Rankin (2004) discovered the double notches in
the leading wing of the main pulse of B0950$+$08.

Navarro et al.~have noticed
that the feature must be `intrinsic to the emission mechanism'
because it becomes narrower at larger observation frequency $\nuobs$.
However, the initial interpretive efforts of theorists did not
follow that suggestion. Wright (2004) 
interpreted the features in terms of
altitude-dependent special-relativistic effects, but was forced
to assume unlikely emission geometry and postulated an opaque
absorber of unknown origin that corotates at/near the light-cylinder. 
This work, however, pioneered the important idea that 
a \emph{single} entity must be responsible for both notches.
Dyks, Fr{\c a}ckowiak, S{\l}owikowska, et al.~(2005) considered
the neutron star itself (embedded in an opaque plasma cloud)
as the absorber/eclipser and their emission region neatly followed
the geometry of magnetic field lines. However,
to make this far more natural geometry really work, the radio emission
had to be directed inward, toward the neutron star.
The model has achieved some agreement with the data, 
but the symmetry of double notches,
as well as their frequency evolution remained unsolved.

\begin{figure}
   \includegraphics[width=0.48\textwidth]{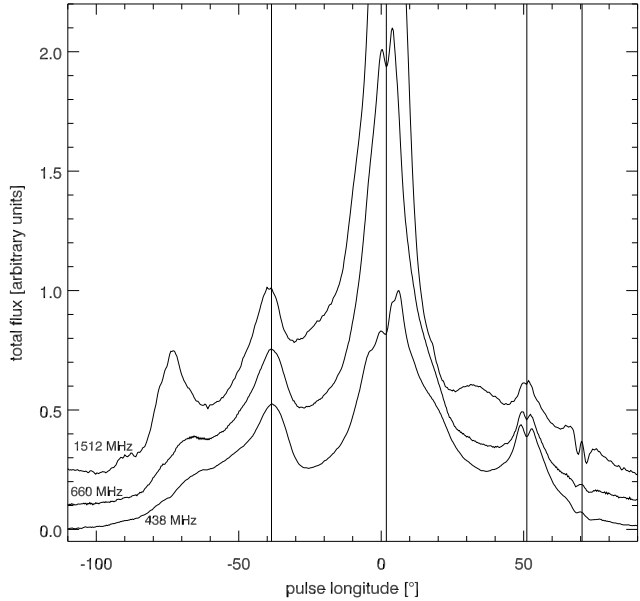}
       \caption{Pulse profiles of PSR J0437$-$4715 at three frequencies.
          The double features at $\phi\simeq 51^\circ$ and $70^\circ$
          make it possible to align the profiles in
          absolute way. Note the lack of ``radius-to-frequency" 
          mapping. Zero levels and flux units are arbitrary and different
          for each profile. Data courtesy: R.~Manchester.
       }
      \label{j04}
\end{figure}

Considerable progress in the ability to interpret  
the data was
done by Dyks, Rudak \& Rankin (2007; hereafter DRR), 
who interpreted the notches
as a direct imprint of the hollow beam intrinsic to the radiation
emitted by electrons accelerated parallel to their velocity.
This work was the first to interpret
the double features in terms of microphysical
beam intrinsic to a specific radiation mechanism.
The model managed to ensure symmetric and double shape 
of the notches whenever
they were not washed out by spatial extent of the emitter/absorber.
The separation $\dn$ of the notches (as observed at that time)
was consistent with the inverse-Compton-like
version of the parallel acceleration maser:
$\nuobs \sim \gamma^2\nupl \propto \dn^{-2}$.
Since the mechanism is intrinsically broad-band,
one could naturally understood the lack of radius-to-frequency mapping 
in J0437$-$4715 (Fig.~\ref{j04}).

DRR noticed that double notches of J0437$-$4715 are located 
in a trailing wing of a bifurcated
emission component (BFC), visible at $\phi\simeq51^\circ$ in Fig.~\ref{j04}.
This component was interpreted
in terms of the same elementary emission beam that produces the 
double notches. Throughout this paper we maintain this view that
the apparently unrelated bifurcated emission components and double notches
have common origin. The notches are regarded as a negative
image of the elementary emission beam of the same radiative process 
that produces the bifurcated components. 

In spite of the overall success of DRR's model, 
the parallel acceleration beam
was unable to explain the observed large depth of double notches.
To produce the notches, a small part of the emitter had to be radio silent
or a localised absorber of unknown origin was needed. Neither it was
possible to decipher the real macroscopic geometry 
of the emitter/absorber system.

In this paper we solve these problems and we identify
both the topology of the elementary emission beam
as well as the general geometry of the eclipsing phenomenon.
This is done by making physical fits to a bifurcated emission component
in the averaged profile of PSR J1012$+$5307 (Section \ref{fitcr}), 
as well as by 3D simulations
of double notches (Section \ref{discr}).
Uncovering of the nature of double features
is equivalent to the identification of the long-sought
radio emission mechanism of pulsars, and results in 
important consequences for several problems of radio pulsar astronomy
(Section \ref{appli}).

\begin{figure}
   \includegraphics[width=0.48\textwidth]{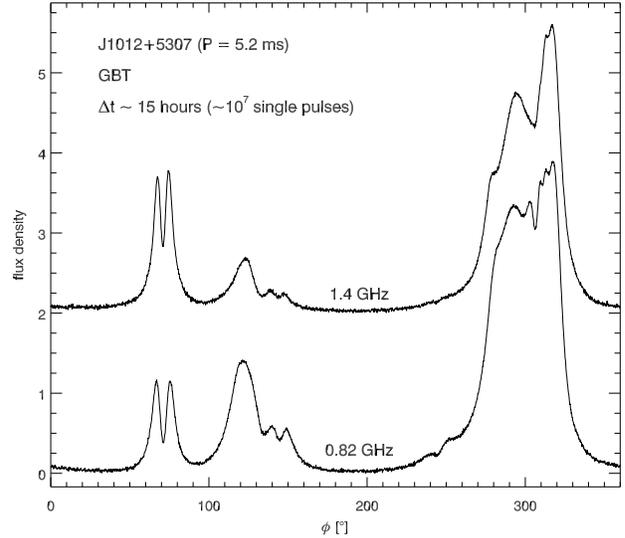}
       \caption{Pulse profiles  of \jtt at two frequencies.
    They are phase-aligned to match the pronounced bifurcated component (BFC) 
    near $\phi\simeq 71^\circ$. This also aligns the central minimum 
    and the trailingmost component of the main pulse (MP). Note the other
 weak BFC located $\sim$$50^\circ$ ahead of the MP in the $0.82$ GHz profile
(around $\phi\sim245^\circ$). 
       }
      \label{jtt}
\end{figure}

\section{The bifurcated component of J1012$+$5307}

The $5.25$-ms pulsar J1012$+$5307 was discovered by Nicastro et al.~(1995).
Its polarisation properties were studied by Xilouris et al.~(1998)
and Stairs et al.~(1999). Kramer et al.~(1999) investigated
the behaviour of this pulsar across the radio frequency spectrum,
and it was the high-quality $1.4$-GHz profile presented in that paper
that has drawn our attention to this object.

Fig.~\ref{jtt} presents the averaged pulse profiles of \jtt
observed with the Green Bank Telescope at $0.82$ (bottom) and $1.4$ GHz 
(top). 
The bandwidth was 64 MHz at both frequencies. The total integration time
of $\sim$$15$ hours ($\sim$$10^7$ single pulses) was accumulated in 
the period between July 2004 and March 2007.
A pronounced, highly symmetric BFC
can be seen near the phase $\phi=70^\circ$
($\sim$$50^\circ$ ahead of the interpulse). Another bifurcated
component (a weak one) is present some $50^\circ$ ahead of the main pulse.

\begin{figure}
   \includegraphics[width=0.48\textwidth]{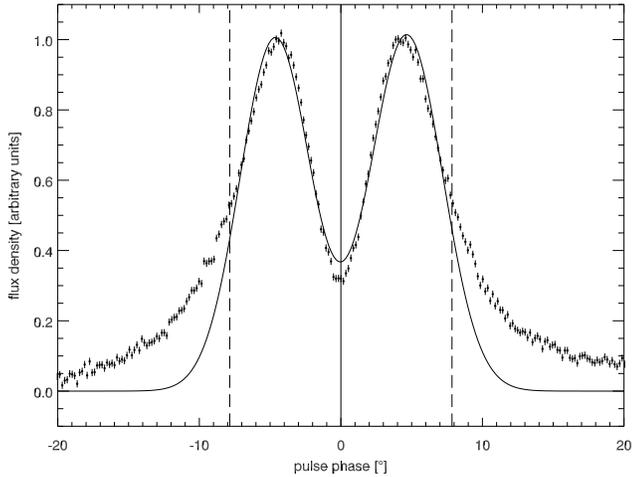}
       \caption{A best fit of a sum of gaussians to the BFC of \jtt\
       at $0.82$ GHz. The vertical dashed lines (also in the next figures)
       mark the phase interval in which the fit was performed.
       In this interval $\chisq\simeq14$.  
       }
      \label{twogauss}
\end{figure}

The outstanding signal-to-noise ratio of the data,
and the strength of the bright double component  make it possible
to test quality of various empirical or physical fits.
In Fig.~\ref{twogauss} we present the traditional decomposition of the BFC
into a sum of two gaussians. The phase interval of the fit
was limited to the central part between the vertical dashed lines.
One can see that modelling of double features with gaussians
is not a good idea. The problem is that any sum of two symmetrical
functions tends to have the inner wings less steep than the outer
wings, which is just opposite to what is observed.\footnote{The 
gaussian fitting is also problematic for single asymmetric components,
as was rightfully emphasized by Weltevrede \& Johnston (2008a).
Still, the authors decided to use symmetric functions.
} 

\section{Physical fits to a radio pulse component}
\label{fits}

Because of astonishing degeneracy of Nature
the BFC component can be fairly well fitted with two different physical
models. To expose the deceiving character of the problem,
we intentionally start the analysis with the incorrect,
parallel-acceleration  model (hereafter PAC model).

\subsection{The wrong idea: parallel acceleration beam} 
 
At $0.82$ GHz, the central minimum in the BFC component of \jtt
reaches a large depth of nearly $70\%$, which in the PAC scenario
could only be interpreted as emission from 
a very localised region. It is therefore assumed that the BFC is free from 
any spatial convolution effects and we model 
it directly with the hollow parallel-acceleration beam.

We use the textbook dipolar radiation pattern for \emph{non-coherent}
radiation emitted by relativistic charges accelerated parallel to 
their velocity (eg.~Rybicki \& Lightman 1979).
The shape of the beam (\emph{frequency-integrated} power per unit solid angle) is 
\begin{equation}
P_\parallel = \frac{q^2}{4\pi c^3}
\ a_\parallel^2
\ \frac{\sin^2\theta}{\left(1 - \beta\cos\theta\right)^5},
\label{pambeam}
\end{equation}
where $\theta$ is the angle between the beam axis (electron velocity)
and the line of sight, $\beta = (1-1/\gamma^2)^{1/2}$ is the electron velocity
in units of the speed of light $c$,
whereas $q$ and $a_\parallel$ are the charge
and acceleration of the electron.

This beam is inclined at some angle $\alpha_b$ with respect to the rotation
axis and our line of sight passes through it at some `beam impact
angle' $\bb = \zeta - \alpha_b$.
The beam intensity \mref{pambeam}
is related to the observed pulse longitude (phase)
through the cosine rule for sides of the spherical triangle
$(\vec \Omega, \vec v, \vec n)$:
\begin{equation}
\cos\theta =
\cos(\phi-\phi_0)\sin\alpha_b\sin\zeta + \cos\alpha_b\cos\zeta,
\label{trigon}
\end{equation}
where $\phi_0$ is the phase of the central minimum in the bifurcated
component, $\vec \Omega$ is the angular rotation velocity of the pulsar,
 $\vec v$ is the velocity of the emitting electron(s), and $\vec n$
is the unit vector of our line of sight. 


The zero-flux level is blindly fixed at the lowest 
place in the profile. Since the inclination of the beam axis
merely rescales the feature by a factor $1/\sin\alpha_b$
we assume $\alpha_b = 90^\circ$.
Thus, the fitted function $F(\phi) = C P_\parallel(\cos\theta(\phi))$
has 4 parameters:
the normalisation constant $C$, the phase location $\phi_0$
of the double component, the viewing angle $\zeta$ (or the beam impact
angle $\bb$), and the electron Lorentz factor $\gamma$.
The fitting was done with the Levenberg-Marquardt method. 
To make sure that the best fit
corresponds to the \emph{global} minimum, the fitting was repeated
for many sets of initial parameters. These
were sampled quasi-randomly from wide intervals, using the maximum-avoidance
algorithm of Sobol', Antonov, and Saleev (Press, et al.~1992).

\begin{figure}
\includegraphics[width=0.48\textwidth]{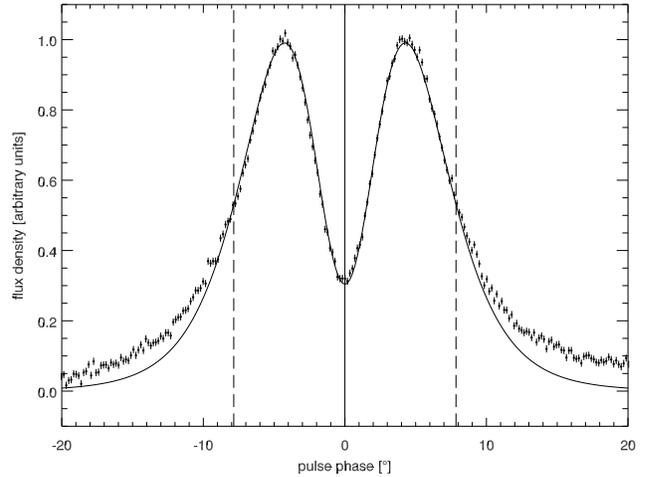}
      \hfill
       \caption{Best fit solution for the off-axis cut through the
frequency-integrated parallel-acceleration beam
 (eq.~\ref{pambeam}). 
Within the marked phase interval $\chi^2/{\rm dof}=2.2$.
       }
      \label{offbeam08}
\end{figure}

The best-fit solution for the $0.82$ GHz profile is shown in
Fig.~\ref{offbeam08}.
The fit was performed for the phase interval
which is marked with the vertical dashed lines
and has a width
of $\Delta\phi=1.8\dbfc$, where $\dbfc$ is the separation
of maxima of BFC.
Within this interval 
$\chi^2/{\rm dof}$ stays close to unity, and it is equal to $2.2$
for the specific case shown in Fig.~\ref{offbeam08}.
The beam impact angle is $\bb = 1.546\pm0.010^\circ$,
and the electron Lorentz factor $\gamma = 6.364\pm0.012$.
For a non-orthogonal $\alpha_b \ne 90^\circ$
this translates to a somewhat larger value of 
$\gamma \simeq 6.4/\sin\alpha_b$.
The maxima are separated by the phase interval 
$\dbfc = 8.54\pm0.020^\circ$ and the opening
angle of the hollow cone is $2\thmax = 9.08\pm0.030^\circ$
(or $9^\circ/\sin\alpha_b$ in the case of $\alpha_b \ne 0$).
The errors include only the statistical $1\sigma$ uncertainty
and are underestimated due to the unknown zero-flux level.
The zero-level is likely to be wrong by few percent 
of the flux
observed at the maxima of the BFC component ($\imax$).
This is suggested by the flux difference of $\sim$$0.01\imax$ observed
at two lowest locations in the $1.4$-GHz profile 
(at $\phi \sim 30^\circ$ and $200^\circ$ in
Fig.~\ref{jtt}). The corresponding systematic errors were estimated
by repeating the fit for various levels of the zero flux. 
A shift by $5\%$ percent of $\imax$ typically
changed the values of $\gamma$, $\thmax$ and $\dbfc$ by $1-2\%$. 
The beam impact angle $\bb$ (and the normalisation constant $C$)
were more sensitive and varied by roughly the same factor ($5\%$).
The actual errors of the parameters are therefore dominated by 
the systematic
effects of unknown zero level and have the magnitude of few percent.
Below, the fitted parameters will be given without errors.

\begin{figure}
\includegraphics[width=0.48\textwidth]{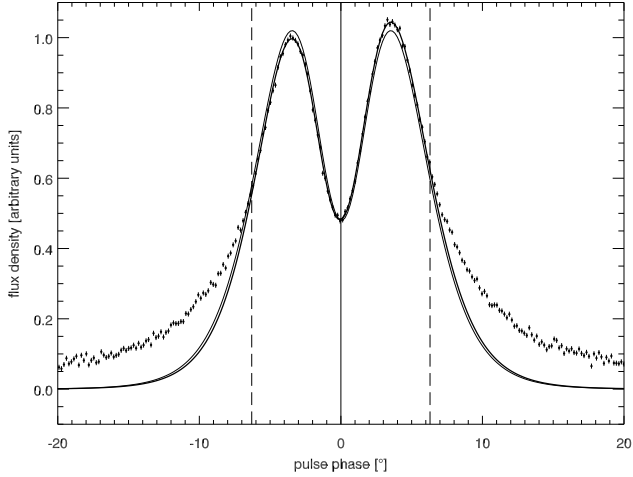}
       \caption{Best fit ($\chisq \simeq 12$, thin line) of the parallel 
acceleration beam to the BFC at $1.4$ GHz. The asymmetric thick line 
presents
a composition of two separate fits to the left- and right-hand part
of the BFC. In this composed case $\chisq \sim 1$.  
       }
      \label{offbeam14}
\end{figure}

In the central parts of the BFC, where bulk of the emission is received,
the parallel acceleration beam reproduces the data with very high accuracy.
This is astonishing, because the fitted function ignores
all spatial and spectral convolution effects
(note that frequency-integrated emissivity is fitted 
to a component observed with fairly narrow bandwidth of
$64/820\ {\rm MHz} = 8\%$). 
The fitted curve also ignores all deformations that could possibly 
arise in the amplification process.

This apparent coincidence is very deceiving and can easily mislead
one to believe in the parallel-acceleration scenario.
However, it merely tells us that the component
can be well modelled with the function of type
$\theta^2/(1 - \beta\cos\theta)^g$, with the exponent $g$
weakly constrained, because it mostly determines the shape of the outer
wings. The PAC is not the only mechanism
described by such function.
So the result does \emph{not} prove the parallel acceleration origin.
It rather suggests that the amplification process
preserves the shape of the non-coherent radiation beam.

\begin{figure}
  \includegraphics[width=0.48\textwidth]{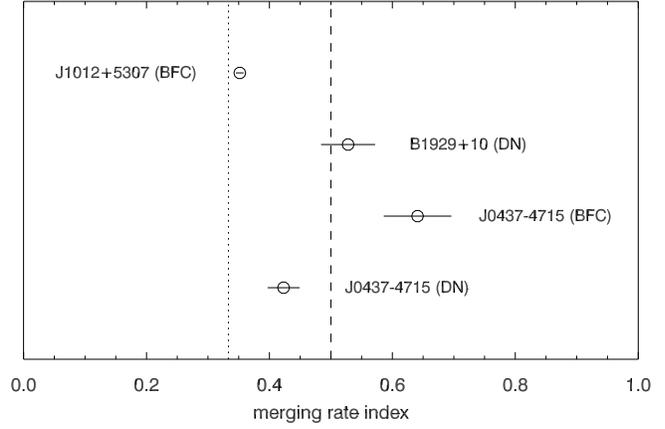}
   \caption{Merging-rate index $a$ in the relations: 
            $\dbfc \propto \nuobs^{-a}$ and $\dn \propto \nuobs^{-a}$, 
    where $\dbfc$ is the separation of maxima in bifurcated components
    (BFC), whereas $\dn$
    is the separation of minima of double notches (DN).
    The vertical lines mark the values of $a$ for the curvature
radiation (dotted) and inverse-Compton-like emission due to the 
parallel acceleration (dashed).
    The errors are statistical $1\sigma$ only. Note that the only
well-resolved and well-modelled case is the BFC of \jttns.
       }
   \label{exponents}
\end{figure}

The fit is worse at $1.4$ GHz (thin line in Fig.~\ref{offbeam14}), 
where the BFC is more asymmetric. However, if the left- and right-hand
half of the BFC is fitted separately (thick line), the fit quality is similar
($\chisq \sim 1$ again within $\Delta\phi \simeq 1.8\dbfc$).
The parameters obtained with the single-step fit are:
$\bb = 1.7^\circ$, $\gamma=7.48$, $2\thmax=7.7^\circ$, $\dbfc = 
6.9^\circ$. The result again looks misleadingly encouraging:
the beam impact angle is, within the few percent errors, 
the same as at $0.82$ GHz. Thus, using the parameters obtained
from the fit at $0.82$-GHz,
it is enough to change only a single parameter
(the Lorentz factor $\gamma$) to reproduce 
\emph{both} the relative depth of the central minimum \emph{and} the
separation of maxima at $1.4$ GHz. 
However, the ratio of $\gamma_{1.4}/\gamma_{0.82}
=1.17$,
that is required to achieve this, is smaller than expected
for the inverse-Compton-like process: the inverse-Compton
relation $\nuobs\propto\gamma^2$ implies $\gamma_{1.4}/\gamma_{0.82}
=1.3$. 
In other words, the inverse-Compton scenario requires that the 
beam size should decrease with increasing frequency as $\thmax\propto
\nuobs^{-0.5}$ whereas the inferred rate is $\thmax\propto\nuobs^{-0.31}$. 
Thus, with the inverse-Compton-implied ratio of $\gamma_{1.4}/\gamma_{0.82}
=1.3$, one cannot fit the data simultaneously at two frequencies.
Instead, the separation of maxima in the 
BFC follows the relation $\dbfc \propto \nuobs^{-0.35}$
(see Fig.~\ref{exponents}) which is fully consistent 
with the low-frequency curvature emission for which 
the opening angle is $\psimax \propto \nuobs^{-1/3}$
(eg.~Jackson 1975).

A much more serious problem with the PAC model is that 
the intrinsic topology of the beam (axially-symmetric hollow cone)
makes it extremely difficult to obtain double notches with
the observed depth of $\ga$$20\%$ (Perry \& Lyne 1985; 
see also fig.~2c in DRR).

\subsection{The curvature radiation}
\label{fitcr}

\subsubsection{The classical radiation beam}

\begin{figure}
   \includegraphics[width=0.48\textwidth]{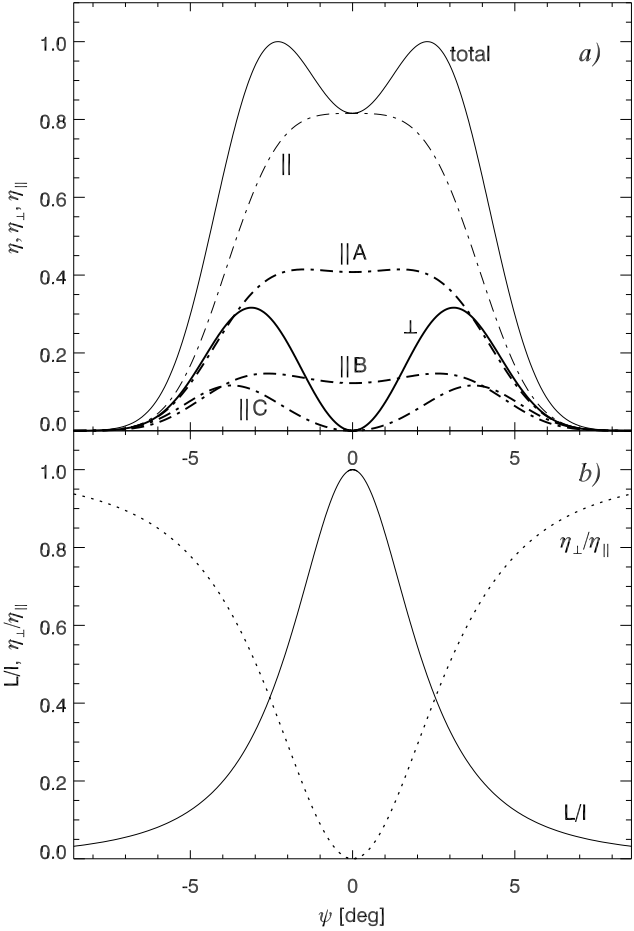}
       \caption{{\bf a)} Transverse crossection through
the fan beam of low-frequency curvature radiation (eq.~\ref{crv}). 
The angle $\psi$ is measured from the \emph{plane} $\psi=0$ 
of electron trajectory (the beam does not have axial symmetry).
The total emissivity $\eta$ (thin solid) consists of the \omode\ radiation
that peaks in the trajectory plane
(thin dot-dashed line) and the \emph{two-planar} \emode\ which 
has the double-peaked crossection shown with the thick solid line.
The thick dot-dashed lines (marked $\parallel$A, $\parallel$B, and
$\parallel$C) show possible profiles of partially-attenuated
\omode\ radiation. {\bf b)} The polarisation degree $L/I$ and the ratio
$\eta_\perp/\eta_\parallel$ for the case with no attenuation of the \omode.
Note the intrinsic decrease of $L/I$ in the outer wings of the fan beam
(see discussion in Sections \ref{polstruc} and \ref{depolsec}).
       }
      \label{etaperp}
\end{figure}

In the case of curvature radiation, the energy emitted per
unit frequency and unit solid angle is given by:
\begin{eqnarray}
\eta_{\rm crv} & = & \eta_\parallel + \eta_\perp =\\
&=& \frac{q^2\omega^2}{3\pi^2c}
\left(\frac{\rho}{c}\right)^2
\left[\xi^2 K^2_{\frac{2}{3}}(y)
+\xi\ K^2_{\frac{1}{3}}(y)\sin^2\psi\right],
\label{crv}
\end{eqnarray}
where
\begin{equation}
\xi = 1/\gamma^2 + \psi^2,\ \ {\rm and}
\label{ksi}
\end{equation}
\begin{equation}
y = \frac{\omega\rho}{3c}\xi^{3/2}
\label{yy}
\end{equation}
(eg.~Jackson 1975).
The symbol $\rho$ denotes the radius of curvature of electron trajectory
in our reference frame,
$\omega = 2\pi\nuobs$, $c$ is the speed of light, and $K$'s are modified
Bessel functions. The angle $\psi$ is measured between our line of sight
and \emph{the plane of} electron 
trajectory.\footnote{The angle $\psi$
should not be mistaken
with the polar angle $\theta$ measured from the electron velocity
or $\vec B$.} The ``well-known" shape of the beam 
is shown in Fig.~\ref{etaperp}.
The beam can be decomposed into two, orthogonally polarised
sub-parts: the
one which is polarised parallel to the projection of electron trajectory
on the sky ($\eta_\parallel$) 
and the other one ($\eta_\perp$), with the polarisation 
orthogonal to the plane 
of the electron trajectory (hereafter ET plane).\footnote{There can be 
a big difference between the shape of electron 
trajectory in the observer's frame and the shape of the $B$-field
line along which the electron propagates. However, it is not important for
the considerations of the present paper, and will be ignored in what
follows. Thus, the ET plane can be considered to be the plane of the 
$B$-field line along which the electron propagates in the corotating frame.}

As shown in Fig.~\ref{etaperp}
all the radiation in the plane of electron trajectory 
($\psi = 0$) is polarised
parallel to the observed projection of the $B$-field line
and is therefore likely to have problems with leaving the magnetospheric 
plasma.
As can be seen in eq.~\mref{crv}, the remaining part of the beam
(the orthogonal mode, hereafter \emode) has the central minimum of exactly the
same shape (as a function of $\psi$)
as the PAC beam has as a function of $\theta$
(near the $B$-field line we have 
$P_\parallel \propto \sin^2\theta$ for the PAC model,
and $\eta_\perp \propto \sin^2\psi$ for the \emode\ curvature radiation). 
The double shape of the orthogonal-mode beam
makes it a suitable tool to model the bifurcated components.

\subsubsection{Fitting the curvature beam}


We assume that the emitter has a form of a thin and elongated
plasma stream that emits the curvature radiation mainly 
in the extraordinary
(orthogonal) mode. At \emph{each} altitude within
some finite but non-negligible range of $\Delta r$, 
a broad range of frequencies is emitted.
Therefore, the beam observed at a fixed frequency
has a fan-like shape and subtends
a range of magnetic colatitudes $\theta_m$. Since there is no 
\emode\ emission in the plane of the stream,
the BFC component is 
naturally produced when our line of sight passes through the plane
of the stream.

We proceed with the fitting as in the parallel acceleration case, 
ie.~we use the \emph{frequency-integrated} power emitted per steradian.
Integrating eq.~\mref{crv}, and dividing by $2\pi\rho/(c\beta)$ one obtains: 
\begin{eqnarray}
I_{\rm crv} & = & I_{{\rm crv,}\parallel} + 
I_{{\rm crv,}\perp} = \\
& = &\frac{7q^2c}{256 \pi \rho^2}
\left(\frac{2\beta^7}{\cos\psi}\right)^{1/2}\times \nonumber\\
& \times & (1 - \beta\cos\psi)^{-5/2}
\left(h(\psi) + \frac{5}{14}
\frac{\beta\cos\psi\sin^2\psi}{1 - \beta\cos\psi}\right),
\label{crvfint}
\end{eqnarray}
(eg.~Konopinski 1981). The function $h(\psi)$ describes the unknown
contribution
of the ordinary (parallel) mode and is equal to $1$ 
in the absence of absorption.
For a pure orthogonal mode, ie.~for $I_{\rm crv} = I_{{\rm
crv,}\perp}$, we have $h(\psi)=0$.
 
To record the bifurcated component, the line of sight must pass through 
the plane of electron trajectory. In the case of the pure orthogonal mode
this implies zero flux at the center of BFC. In reality, however, 
there are several factors that are likely to raise the central minimum.
These include a small (but non-zero) thickness of the plasma stream,
partial admixture of the  \omode, and non-planarity of the
trajectory. To be able to fit the central part of the BFC we 
artificially raise the level of the central minimum in two ways:
1) $h(\psi)$ is considered as a $\psi$-independent
normalisation parameter for the parallel mode ($h(\psi)=h_0 < 1$); 
2) we redefine the
meaning of the angle $\psi$.

For numerical simplicity, we assume that the plane of the stream is 
meridional, and the general geometry of the problem is
quasi-orthogonal: $\alpha_b=90^\circ$, $\zeta \sim 90^\circ$.
In such a case $\psi$ simply becomes
\begin{equation}
\psi \simeq \phi-\phi_0,
\label{psireal}
\end{equation}
where $\phi$ is the observed pulse phase and $\phi_0$ is the phase at which
the center of the BFC is observed. So the situation is different
than in the case of eq.~\mref{trigon}. 
However, it is convenient to mimic the non zero flux 
at the center of the profile by assuming 
$\cos\psi \simeq \cos\theta$ as given by 
eq.~\mref{trigon}. This equation reduces to eq.~\mref{psireal}
whenever $\alpha_b-\zeta \ll \phi-\phi_0$, ie.~everywhere except
near the central minimum, since in our case $\alpha_b -\zeta 
\simeq 1.65^\circ$.
The value of $\zeta$ fitted in that way has no physical or geometrical
meaning. It is an arbitrary parameter that allows to mimic 
the unknown contribution of the ordinary-mode radiation
(or other effects).

Thus, the fitted function was either given by eq.~\mref{crvfint}
with $\psi=\phi-\phi_0$, or by the following formula:
\begin{equation}
\iperp = A\ \frac{\cos^{1/2}\psi\ \sin^2\psi}{(1 - \beta\cos\psi)^{7/2}}
\label{simplecr}
\end{equation} 
with $\cos\psi=\cos\theta$ given by eq.~\mref{trigon}. Both methods
gave similar parameter values and had comparable $\chisq$,
with eq.~\mref{simplecr} doing only slightly better than \mref{crvfint}.

\begin{figure}
   \includegraphics[width=0.48\textwidth]{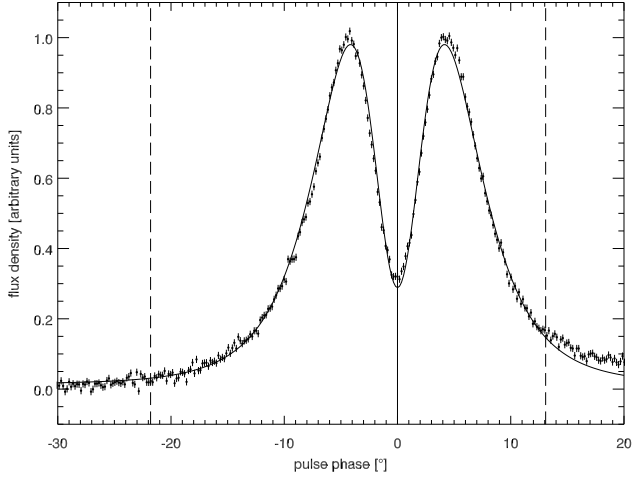}
       \caption{Best fit of the frequency-integrated and \emode-dominated
curvature beam (eq.~\ref{simplecr}) to the $0.82$-GHz BFC
of \jttns. Note the much wider interval of the fit, in which
$\chisq \simeq 3.6$.
       }
      \label{crvpheno08}
\end{figure}

\begin{figure}
   \includegraphics[width=0.48\textwidth]{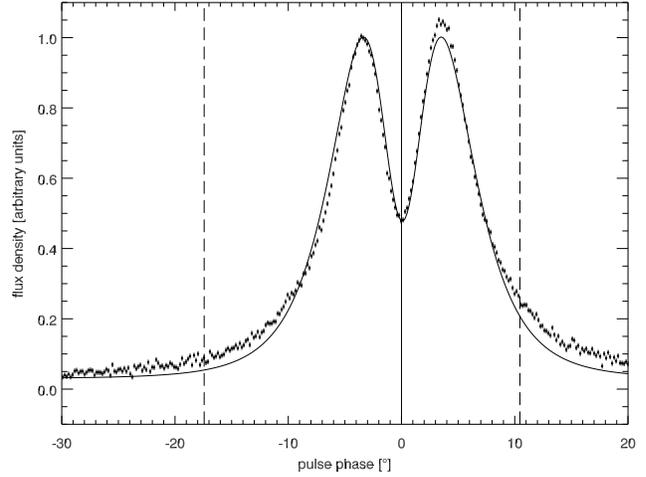}
       \caption{The curvature-radiation fit at $1.4$ GHz. 
       }
      \label{crvpheno14}
\end{figure}

Fig.~\mref{crvpheno08} shows the fit of eq.~\mref{simplecr}
to the $0.82$ MHz profile.
The agreement is only slightly worse than in the case of the parallel
acceleration beam ($\chisq = 3.6$) but holds within a much larger
interval of phase (marked with the vertical dashed lines). 
A fit at $1.4$ GHz (Fig.~\ref{crvpheno14})
is worse due to the BFC's asymmetry and likely due to increased
contribution of the \omode.

The extraordinary curvature beam works better in the outer wings
because of the much smaller exponent ($g=7/2$) in the denominator of 
eq.~\mref{simplecr} (cf.~eq.~\ref{pambeam} in which $g=5$). [Note: Because
in the outer wings we have $1-\beta\cos\psi \simeq \psi^2/2$,
the wings have the same shape of $(1-\beta\cos\psi)^{-5/2}$ for
both the parallel and orthogonal polarisation state
(see eq.~\ref{crvfint}).
Therefore,
we propose that single pulse components be fitted with the function of type:
\begin{eqnarray}
I(\phi) & = & A(1-\beta\cos(\phi-\phi_0))^{-5/2}\simeq \\
& \simeq & 
A^\prime\left(\frac{1}{\gamma^2} + (\phi-\phi_0)^2\right)^{-5/2},
\label{profit}
\end{eqnarray}
where $\beta=(1-1/\gamma^2)^{1/2}$, and $\gamma$, $A$, and $\phi_0$ are the
parameters of the fit. The value of $\gamma$, which determines the
steepness of wings should be different for the leading and trailing half
of an asymmetric component. Such function is very successful
in reproducing the wings of profiles and allows to avoid the false
components that usually come out in the procedure of gaussian fitting
(see fig.~2 in Weltevrede \& Johnston 2008a and the comments therein).]

However, 
the agreement visible in Fig.~\ref{crvpheno08} is not ideal and the fitted function
again ignores the issue of limited bandwidth as well as all
possible convolution effects. Therefore,
the quality of the fits shown in Figs.~\ref{offbeam08} and \ref{crvpheno08},
if considered alone, does not allow us to firmly
discriminate between the PAC maser and the curvature radiation.
What indicates the curvature origin of the beam, is that
the fitted separation of peaks ($8.30^\circ$ and $6.87^\circ$
at $0.82$ and $1.4$ GHz)
is consistent with the relation $\dbfc \propto \nuobs^{-1/3}$.
Such relation is expected 
if the observed frequency
is smaller than, or equal to the
characteristic frequency $\nucr$ of the curvature spectrum: 
\begin{eqnarray}
\nucr = \frac{3c}{4\pi}\frac{\gamma^3}{\rho}  =   
7\negthinspace\cdot\negthinspace10^9\thinspace{\rm Hz} 
\frac{\gamma^3}{\rho/(1\thinspace {\rm cm})}
 =  0.7\thinspace{\rm MHz}\frac{(\gamma/10)^3}{\rho/(10^7\thinspace {\rm cm})}.
\label{nucr}
\end{eqnarray}
In both these cases ($\nuobs \ll \nucr$, or
$\nuobs \simeq \nucr$)   
the opening angle of the beam is $\psimax \sim 
[3c/(2\pi\rho\nuobs)]^{1/3}$, which
follows directly from the properties of the fixed-frequency radiation
pattern (eq.~\ref{crv}, eg.~Jackson 1975).
The exponent of $1/3$ matches
the behaviour of the BFC of \jttns: the observed $\dbfc$ can be well-fitted
\emph{simultaneously} at two frequencies, with the ratio of
$\gamma_{1.4}/\gamma_{0.82}$ fixed at the predicted value of 
$(1400/820)^{1/3}=
1.195$. However, the exponent of $1/3$ is 
not ubiquitous among pulsars
(see Fig.~\ref{exponents}). An additional criterion is evidently needed to
discriminate the models.

\section{Depth of notches as The Decisive Discriminator}
\label{discr}

\begin{figure}
\includegraphics[width=0.49\textwidth]{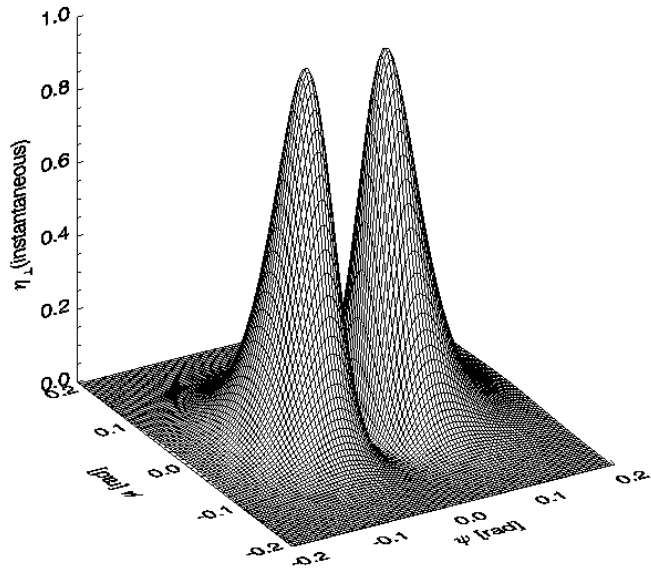}
       \caption{Instantaneous beam of \emode\ curvature radiation
(power per unit frequency per unit solid angle) as a function of
the angle $\psi$, measured from the plane of a $B$-field line,
and the angle $\mu$ measured in the orthogonal way.
The vector of electron velocity (not shown)
is anchored at the origin of the
reference frame and points vertically upward, along the local magnetic field. 
There is no radiation
in the plane of the electron trajectory, which is the plane of $\psi=0$.
       }
      \label{beam3D}
\end{figure}

As shown in DRR the PAC maser is unable to naturally produce
double notches with the observed depth of $\ga20\%$ (Perry \& Lyne 1985). 
This is because
the PAC beam has the hollow cone shape, and is axially symmetric
 around the local electron velocity
(around the local $\vec B$). Therefore, a ring-shaped part of the emitter
contributes to the flux that is observed at the minima of notches
and makes them very shallow.
For a two-dimensional emitter that extends laterally
at a fixed altitude, this `light-polluting' 
ring
has the crossection which
is marked with letters `b' in fig.~3 of DRR.

\begin{figure}
   \includegraphics[width=0.47\textwidth]{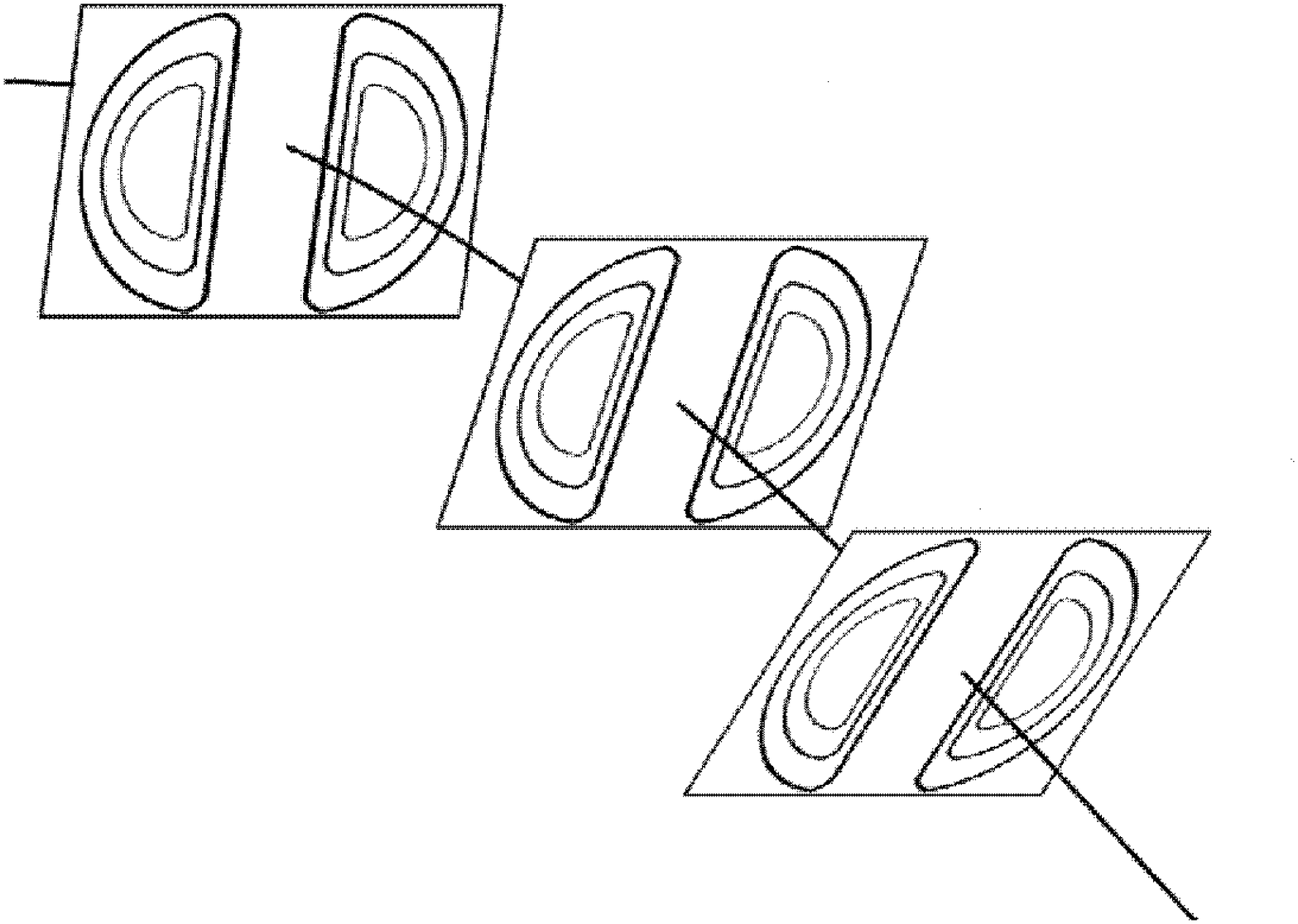}
       \caption{The beam of the previous figure is carried by the
electron along a magnetic field line, which produces
the two-planar, split-fan beam. 
The fixed-intensity contours of the instantaneous beam are shown 
at three locations on a magnetic field line.}
      \label{deski}
\end{figure}

\begin{figure}
   \includegraphics[width=0.49\textwidth]{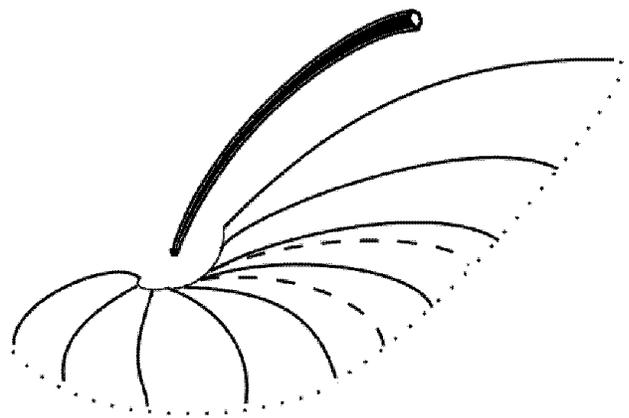}
       \caption{Natural geometry that produces deep double notches.
    An opaque stream (black arc) is located above a two-dimensional
  emission region. The solid lines present magnetic field lines.
 Two dashed lines trace the eclipsed parts of the emitter.  
       }
      \label{parasol}
\end{figure}

In the case of the \emode\ curvature radiation, 
the beam has the mirror symmetry with respect to the plane of the 
local $B$-field line, and there is no (or little) emission within the entire 
plane (see Figs.~\ref{beam3D} and \ref{deski}).
Each point of emission region is therefore 
mostly emitting in only two directions, which make small angles
$\psimax$ with respect to the plane of the $B$-field line
(note that in the case of the low-frequency curvature radiation
 $\psimax\gg1/\gamma$).
Therefore, instead of the `polluting ring', there are only two spots
in the emission region (on both sides of the absorber)
that contribute to the flux observed at the minima of double
notches. The minima are therefore
much less contaminated by the nearby parts of the emitter
than in the PAC case. Moreover, the eclipsing object does not longer
need to be compact.
Overlying opaque streams (Fig.~\ref{parasol}), or
elongated disruptions
in the emitter, can produce
deep notches without affecting the flux at the center of the notches. 
In the curvature scenario deep 
double notches are therefore produced easily and naturally.\footnote{After 
many failed numerical attempts to reproduce the observed
depth of notches (using the axially-symmetric beam of PAC), 
DRR concluded: ``The large depth of double notches
could easily be produced if the coherent radio emission had occured at 
two small angles ($1/\gamma$) with respect to the plane of B-field
lines."}

Fig.~\ref{deski} schematically shows how the split-fan beam 
due to a thin stream is created:
when electrons move along the curved magnetic field lines
they carry-along the two-directional pattern of Fig.~\ref{beam3D}.
The resulting beam has therefore completely different structure
(split fan) than the parallel-acceleration beam (hollow cone).
It is precisely this topological property 
that really makes the crucial difference, and allows us to recognize which
mechanism is the correct one.

The ease with which the \emode\ curvature beam produces deep double notches,
was verified with a 3D numerical simulation. The code assumes that
the emitter has the shape of a surface formed by magnetic
field lines with fixed footprint parameter $s$.
The footprint parameter is arbitrarily fixed at $s=1$ (last open field
lines). To produce the omnipresent `pedestal' emission,
the curvature radiation continues up to a quite high altitude
of $(2/3)\rlc$.\footnote{The pedestal can also be created
by \emph{low-altitude} emission from $B$-field lines 
with $s>1$ (closed field line region) or even from fixed-altitude region
involving a range of $s$.}

The most natural way to generate the notches is shown in Fig.~\ref{parasol}
and involves a thin
opaque stream of plasma above our emission region. We call it most natural
because we know
that such streams exist and produce the BFC.
Note that because the notches are observed in \emph{averaged} profiles,
the obscuring stream has to be fairly permanent to be present in the same
pulses as the emission which is absorbed.
However, to simplify the calculations we preferred to model an elongated
fissure in the emitter, instead of the obscuring stream.
Thus, we assumed that a narrow
part of the emitter (a wedge with a fixed angular width
of $\Delta\phi_m=4.1^\circ$ in the magnetic azimuth)
is not emitting. 
This is a safe assumption, since the 
ability to produce deep notches
does not depend on whether they are produced
by an absorber above, or the hole within the emission region
(DRR).

\begin{figure}
\includegraphics[width=0.5\textwidth]{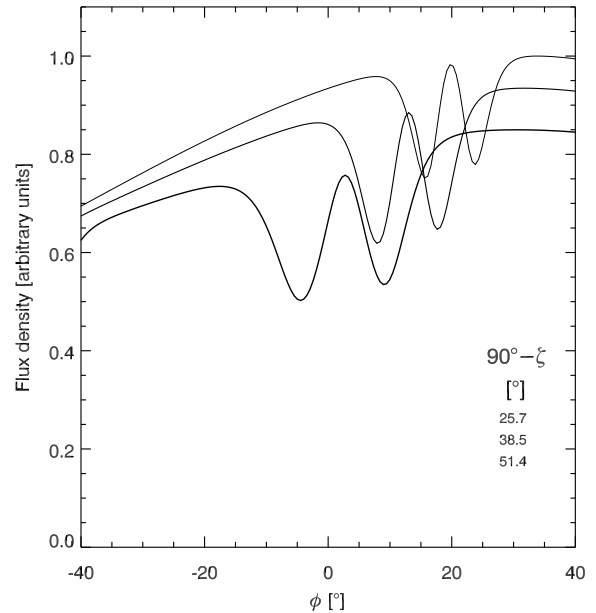}
       \caption{Deep double notches simulated with a 3D numerical code.
      The background (pedestal) emission was created
      by the surface of last open field lines.
A wedge-shaped part of the surface
was not emitting to produce the notches. The wedge was centered at the 
magnetic meridian $\phi_m=-20^\circ$, and had the azimuthal width of
$\Delta\phi_m = 4.11^\circ$.
The curves correspond to $\zeta = 64.3^\circ$, $51.5^\circ$, and $38.6^\circ$ 
(top to bottom). The magnetic dipolar field was tilted 
at the angle $\alpha=90^\circ$.
       }
      \label{kielich}
\end{figure}

The simulations have confirmed that deep double notches
appear naturally for the \emode\ curvature beam. A sample result
for three different viewing angles 
is shown in Fig.~\ref{kielich}. It also becomes clear now why the notches
can only be observed in \emph{very highly polarised} pedestal
emission: when the \omode\ is present, it fills-in the central minimum
of the double \emode\ beam and the notches cannot appear.

Based on Sections 3 and 4 we learn that in the case of 
curvature radiation, 
both the double absorption features (double notches)
of several pulsars
as well as the bifurcated emission components of \jtt and J0437$-$4715
can be understood in terms of natural geometry 
that involves the absorption/emission by plasma
streams in pulsar magnetic field.
We conclude that 
it is the curvature radiation, not the radiation due to the 
parallel-acceleration,
that is observed as the coherent radio emission from pulsars.

\section{Polarisation structure of the curvature beam}
\label{polstruc}

To consider some implications of our findings,
we first discuss the polarisation structure of the
two-modal curvature beam.

It is convenient to imagine the `elementary' beam 
as the fan beam due to the passage of an electron
along a magnetic field line
(or due to a thin stream flowing along a narrow bunch of $B$-field lines).
Based on the fits of Section \ref{fits} it is reasonable
to assume that the $\perp$-mode
beam is amplified isotropically and preserves 
the original (non-coherent shape). The $\parallel$-mode
beam is likely to be attenuated in the central parts
(near $\psi \sim 0$), where the radiation propagates through
the largest plasma density.
The outer parts of the $\parallel$-beam are emitted from the outer boundary
of the stream
into the low-density region surrounding the stream.
Guided by the observed edge depolarisation of profiles
(see Section \ref{depolsec})
we assume that 
in the outer wings of the beam, the \omode\ radiation
is amplified with the same efficiency as the \emode\
and is not attenuated.

The polarisation-structure of such a toy beam is shown in Fig.~\ref{etaperp}a
with thick lines (solid for \emode, dot-dashed for \omode).
Three cases of \omode\ attenuation are shown, marked with $\parallel$A,
$\parallel$B, and $\parallel$C. Since the \emode\ has zero emissivity 
in the ET plane, even a small contribution of \omode\ 
(shown as the case $\parallel$B) can dominate the central parts of the
total beam. It is also possible to have the \omode\ fully absorbed
in the beam center (case $\parallel$C).
Such half-phenomenological curvature-radiation beam has several intrinsic
properties that make it notably successful in interpreting
enigmatic pulsar phenomena.\footnote{Progress in understanding of 
the polarisation structure could be achieved by decomposing the BFC of \jtt
into orthogonal polarisation modes. Unfortunately, the polarisation data
for \jtt are not available to us at the present moment.}

\section{Application to selected pulsar problems}
\label{appli}

\subsection{Polarised multiple imaging}

The inferred polar carousel of many pulsars exhibits interesting structure
in magnetic azimuth $\phi_m$: subbeams of opposite orthogonal 
polarisation are distributed alternatingly along the polar ring of emission.
This pattern produces several enigmatic phenomena, such as
1) the ``modal parity at the outside edges of polar beam" (Rankin
\& Ramachandran 2003), 2) two orthogonally-polarised, out-of-phase drift
patterns, 3) jumps in drift phase (Edwards 2004).
The latter two effects were described by
Edwards, Stappers \& van Leeuwen (2003) 
as `multiple imaging' possibly caused by refraction, birefringence
or aberration/retardation effects.

Without intending to model the phenomena in detail, 
we want to show that the low-frequency curvature radiation beam
has several intrinsic properties that are consistent
with the observations.
First, the multiple imaging is automatically implied by the non-zero 
angular width
of the low-frequency beam of curvature radiation:
\begin{equation}
\psimax \simeq 0.62\left(\frac{3c}{2\pi\rho\nuobs}\right)^{1/3}
=0.40^\circ (\rho_7 \nu_9)^{-1/3},
\label{crwidth}
\end{equation}
where $\rho_7 =\rho/(10^7$ cm), and $\nu_9 = \nuobs/(1\ {\rm GHz})$. 
One can see that for the typical radius of curvature
of $10^7$ cm, the intrinsic width effects
become important/noticeable right in the lower part of the standard
frequency band of radio astronomy ($\nuobs < 1$ GHz). Any intrinsic
broadening of components that is smaller than $\sim$$1^\circ$ would be hard
to notice in pulse profiles.
Thus, by transiting from
sub-GHz to super-GHz radio band one goes from profiles dominated by
intrinsic shape of the curvature-radiation beam 
(with the radiation emitted at noticeable angles with respect to $\vec B$)
to profiles
dominated by the spatial distribution of emission region
(with the emission tightly bound to the direction of $\vec B$).
At fixed $\nuobs$ 
the intrinsic broadening 
depends only on the curvature radius $\rho$ and the dependence
is rather weak ($\propto\rho^{1/3}$). 
This explains why
the scale of double features is so similar in so different objects like
the millisecond pulsars (MSPs) and normal pulsars.
On the other hand, $\rho$ is a bit smaller in MSPs
and the strongly flaring $B$-field lines make it possible
to observe the double emission components well-isolated from the main pulse. 
Therefore,
the double emission features are most easily noticed
in MSPs. The scale of double absorption features
(notches) additionally depends on the eclipsing geometry and will be
analysed elsewhere.
We conclude that the multiple imaging effect
is naturally guaranteed by the non-zero width of the
curvature beam at low $\nuobs$. No refraction, birefringence
etc.~is needed in the zeroth-order approach to understand the 
multiple imaging.

\subsubsection{A method to determine the curvature
of electron trajectory in the emission region}

At a fixed and low observation frequency ($\nuobs \le \nucr$),
the opening angle $\psimax$ of curvature emission
depends only on the curvature radius of electron trajectory
in our (observer's) reference frame.
This opens a way to calculate the curvature radius 
$\rho$ in the emission region from eq.~\mref{crwidth}. 
However, this can only be achieved when full information 
on the global geometry of the system is available. 
Since the viewing angle $\zeta$ can be different than $90^\circ$,
the observed separation of maxima in the BFC can be rescaled
(increased) by the factor $1/\sin\zeta$ due to the well-known
`not-a-great-circle' effect.
Moreover, if the plane of the electron trajectory is inclined
at an angle $\dmdn$ with respect to the local meridional 
plane\footnote{The rotational meridian at which our line of sight crossess
the plane of electron trajectory is meant here.}
the value of $\dbfc$ is increased by additional factor $1/\cos\dmdn$.
One can also use the factor $1/\sin\bfb$, where $\bfb=90^\circ-\dmdn$ 
is the angle at which we cut the ET plane.\footnote{Thus, $\bfb$ is the angle 
between the ET plane and the plane that is locally tangent to the surface of
the cone encircled by our line of sight.}
The observed $\dbfc$ is therefore related to $\psimax$ through
the global geometry parameters:
\begin{equation}
\dbfc \simeq \frac{2\psimax}{\sin\zeta\cos\dmdn}=
\frac{0.8^\circ}{\sin\zeta\sin\bfb(\rho_7\nu_9)^{1/3}}.
\label{dbfc}
\end{equation}
Therefore:
\begin{equation}
\rho_6 = \left(\frac{1.72^\circ}{\dbfc}\frac{1}
{\sin\zeta\sin\bfb}\right)^3 \nu_9^{-1},
\label{rho}
\end{equation}
where $\rho_6=\rho/(10^6\ {\rm cm})$.
For \jtt we have $\dbfc = 8.61^\circ$ and $7.1^\circ$
at $0.82$ and $1.4$ GHz respectively, which implies
$\rho\sin^3\zeta\sin^3\bfb \simeq 0.011\cdot 10^6$ cm at 
both frequencies.
To make $\rho$ equal to $10^6$ cm, either $\zeta$ or $\bfb$ of $\sim$$13^\circ$ 
is required.
For moderate values of $\zeta=\bfb\simeq45^\circ$, the value of $\rho$
is increased by a factor of $8$ up to $\sim$$0.1\cdot 10^6$ cm.

Note that the small value of $\rho\sin^3\zeta\sin^3\bfb$ does not have to 
necessarily imply sub-stellar curvature radius.
The main pulse of \jtt has the width that corresponds exactly to the opening
angle of low-altitude polar tube, and the interpulse is present.
This suggests orthogonal viewing geometry 
($\zeta \sim \alpha  \sim 90^\circ$). 
However, this kind of geometry simultaneously necessitates
small cut angles
($\bfb \sim 0^\circ$) for components
observed well ahead of main profile features (main pulse and interpulse). 

In the case of J0437$-$4715, we have $\dbfc = 3.76^\circ$, $2.89^\circ$,
and $1.7^\circ$ at $0.438$, $0.66$ and $1.512$ GHz, respectively.
This gives $\rho\sin^3\zeta\sin^3\bfb = 0.22$, $0.3$,
and $0.7\cdot10^6$ cm, respectively. The results for J0437$-$4715
are closer to what we expect for dipolar magnetic field lines.
The actual $\rho$ is probably larger, because $\zeta \la 45^\circ$ 
is likely. On the other hand, the BFC of J0437$-$4715 has 
a very shallow central
minimum due to some convolution effects (eg.~due to the $\vec E\times \vec
B$ drift of the stream). This can make $\dbfc$ appear 
slightly smaller.

It is worth to digress here into the phase location of BFCs 
in pulse profiles. The precursor location of both BFCs observed in \jtt
(at $\phi \simeq 71^\circ$ and $245^\circ$ in Fig.~\ref{jtt})
is reasonable, because
for fast-rotating pulsars with large dipole inclination,
radiation emitted from any altitude cannot be detected 
far on the trailing side of the main pulse.
Because of the effects
of aberration and retardation, the latest-possible
phase of detection is equal to the caustic phase of 
$16^\circ s^2/\sin^2\alpha 
+ \phi_f$,
where $s$ is the footprint parameter of $B$-field lines and $\phi_f$
is the absolute fiducial phase (Dyks, Wright \& Demorest 2009, 
in preparation). J0437$-$4715 does not obey this `precursor rule',
probably because of much smaller dipole inclination ($\alpha \ll 90^\circ$,
$\zeta \ll 90^\circ$).

\begin{figure}
   \includegraphics[width=0.48\textwidth]{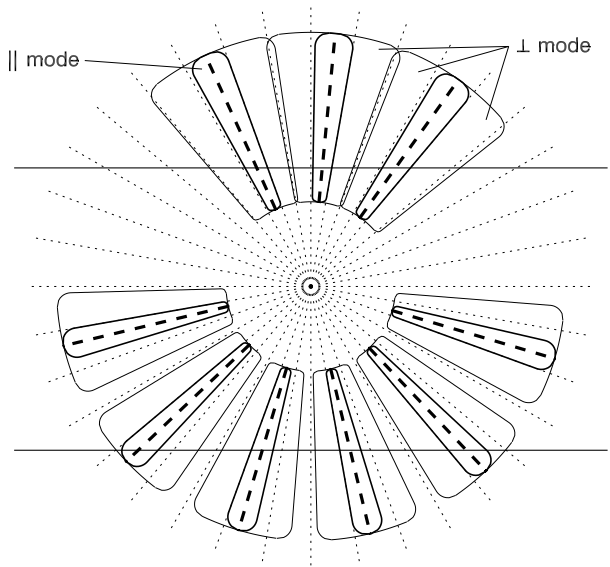}
       \caption{In the case of the curvature radiation with 
                partially-absorbed \omode,
                the carousel of electron streams (thick dashed) 
               naturally produces
               polarisation pattern in which the two orthogonal modes
               dominate alternatingly in magnetic azimuth.
               The thick contours present the \omode, 
               the thin contours are for the \emode, and the $B$-field
               lines are dotted. The two
               horizontal lines mark two different ways in which 
               our line of sight can leave the beam: in the case shown at
               the top the sightline cuts through
               the fixed-azimuth outer boundary 
             of an azimuthally-limited emitter. 
       }
      \label{caro}
\end{figure}

\subsubsection{Azimuthal separation of orthogonal modes}

The plane of $\psi=0$,
(orthogonal to the page in Fig.~\ref{etaperp}a),
is the plane of symmetry of the curvature beam. 
Such \emph{fan} beam has therefore
intrinsic mirror symmetry with respect to the plane of $B$-field lines.
In the case of the partial contribution of the \omode\ 
(case $\parallel$B in Fig.~\ref{etaperp}),
the two orthogonal polarisation modes are naturally separated
in magnetic azimuth: the \omode\ dominates in the plane of the
$B$-field lines, whereas the \emode\ dominates on both sides of the
parallel mode. 
This provides a very natural way to obtain the
azimuthally-structured polarised carousels observed by 
Rankin et al.~(2006) and 
Edwards et al.~(2003).

Let us assume the usual picture in which the polar region
is populated with thin plasma streams that are 
separated in magnetic azimuth and are anchored to the sparking
spots (the carousel of Ruderman \& Sutherland 1975).
In such a case, the geometry of the curvature beam 
contaminated by the \omode\
assumes the shape shown in Fig.~\ref{caro}. The thin plasma streams
(actually only their fragments
that emit detectable radio waves)
are marked with thick dashed lines. 
The \omode\ emission (dominating within, but not constrained to, 
the planes of the streams) is delineated with thick solid contours. 
The \emode\ emission flanks the sides of the \omode\ beam
(thin contours). The pattern naturally results
from the intrinsic properties of the curvature beam with not-fully-absorbed
\omode. Thus, the azimuthal separation of modes can appear even with no
refraction nor birefringence involved.

Note that Fig.~\ref{caro} presents the instantaneous geometry of
emitting streams, and refers to single pulse data after they have been 
deconvolved into the image of polar carousel. In the averaged pulse profile,
the structure of Fig.~\ref{caro} is likely to become smeared out 
by plasma drift effects. The exact outcome of such averaging depends 
on the relative speed of drift motion and pulsar rotation.

\subsection{Edge depolarisation of profiles}
\label{depolsec}

Rankin \& Ramachandran (2003) emphasized the significant observation
that the outer edges of pulsar profiles are almost totally depolarised.
Thus, some high-level conspiracy between the modes seems to be required.
The curvature radiation is an extremely likely cause of this conspiracy,
because it intrinsically emits equal amounts of the orthogonal
polarisation  states in the outer edges of the curvature beam.
For $\psi \gg \psimax$, we have $\xi =1/\gamma^2 +\psi^2 \simeq \psi^2$,
and $K_{\frac{2}{3}}(y) \simeq K_{\frac{1}{3}}(y)$,
so the eq.~\mref{crv} implies:
\begin{equation}
\eta_\perp/\eta_\parallel = 1.
\label{depol}
\end{equation}
This ratio, as well as the polarisation degree $L/I$,
are shown as a function of $\psi$ in Fig.~\ref{etaperp}b.
The decrease of $L/I$ with the angular distance from the symmetry plane
(increasing $|\psi|$) is readily apparent. 

The well-known problem is that the emitted curvature radiation is
polarised elliptically. It is not only limited to the two orthogonal modes.
So some mechanism is needed to filter out the 
purely-orthogonal modes.\footnote{Several observations show that the separation
can be very inefficient, and the observed radiation is also polarised
elliptically (eg.~B2044$+$15, outer wings of B0525$+$21, B0751$+$32
and many other cases, see eg.~Hankins \& Rankin 2008;
Edwards \& Stappers 2004).} 
Whatever mechanism actually does it, a seemingly natural way 
to have equal amounts of two strictly-orthogonal modes, 
is to filter them proportionally out of a beam that already 
(intrinsically) includes
equal amounts of radiation in the two orthogonal polarisation states.
The curvature emission is then an ideal candidate, because
in the outer wings of the elementary beam, 
it intrinsically produces equal amounts
of radiation in the appropriate polarisation states.
We therefore claim that the observed edge depolarisation has its origin in
the intrinsically equal emissivity in both polarisation
states. Our line of sight can pass through 
the pulsar beam shown in Fig.~\ref{caro}
in two ways, marked with the horizontal lines. In the bottom case
(grazing the edge of the carousel) it is probably a bit difficult to guess
the averaged polarisation degree without numerical simulation.
In the top case, however, the polarisation degree at the edge of the profile
will decrease, because our line of sight moves away from the fixed-azimuth
plane of the last-seen emitter ($|\psi|$ in Fig.~\ref{etaperp}b increases).

\subsection{The origin of double conal profiles}

An interesting question is what happens when the stream
has noticeable azimuthal width, that is comparable to (or larger than)
the opening angle $\psimax$.
We argue that a cut through such stream produces the double conal
profiles (type D in the classification scheme of Rankin 1983).
A geometry of such a cut is presented in the bottom right corner
of Fig.~\ref{bcake}. The brightening of the edges of the profile
is caused by the increased contribution of
\omode. The near-absence of the \omode\ in the inner parts
of the profile leads to the smaller flux and large polarisation degree.
The presence of the \omode\ at the flanks of the profile
also produces the depolarisation. In some cases, when sufficiently
large amount of the \omode\ gets through, orthogonal jumps 
of averaged polarisation degree are observed 
in the outer wings of a pulse
(eg.~B1133$+$16, B2020$+$28,
B2053$+$21, Hankins \& Rankin 2008, hereafter HR08).

The nearly-complete depolarisation of profile's edges is ensured by the
fact that our line of sight passes through a region with limited
magnetic azimuth (as can be seen in Fig.~\ref{bcake}
the sightline enters/exits the fixed-azimuth sides of the elongated
wedge). It is not passing through a conal ring centered at the dipole
axis. The brightest parts of the emission region 
(and the peaks of a profile) correspond to 
the fixed value of $\phi_m =\phi_m^{\rm le}$ on the leading edge
and to $\phi_m =\phi_m^{\rm te}$ on the trailing edge of the
wedge-shaped stream. When our line of sight departures from the
fixed-$\phi_m$ edge, it moves away from the center of the curvature beam
shown in Fig.~\ref{etaperp}a. The polarisation degree is therefore
decreasing as shown in Fig.~\ref{etaperp}b.

\subsubsection{The origin of the radius-to-frequency mapping}

The interpretation here implies that the observed pulse width $W$, 
as measured between
the profile peaks is fixed by the extent of the
emitter/stream in the magnetic azimuth: $\Delta\phi_m = \phi_m^{\rm te}
-\phi_m^{\rm le}$. In the zeroth-order consideration this implies
the lack of pulse broadening at low frequencies.\footnote{In this paper
we reserve the name ``radius-to-frequency mapping" (RFM) only
to the geometric \emph{interpretation} of the broadening. 
It is not used in reference to the \emph{observed phenomenon}.} 
Some of the D-type pulsars indeed exhibit no widening within
a huge frequency range (eg.~B0834$+$06 between $\sim25$ and $2370$ MHz, 
see Fig.~4 in HR08), but some definitely do (eg.~B0301$+$19). 

However, on a closer inspection
the widening (mis-interpreted as RFM) can appear at low frequencies, 
since the plasma density in the
stream falls off with the transverse distance from the stream center/axis.
So we assume that the basic principle of the broadening is the same as usual:
low-$\nuobs$ radiation emerges from regions with lower density.
However, the lower density regions in our scenario are simply
the more extreme outskirts of the stream, ie.~the lower-density
region located further away from the stream axis.
Thus, we replace the `radius-to-frequency' mapping
with the `stream-diameter-to-frequency' mapping.
This interpretation implies that all radio frequencies come 
from the same (or similar) range of altitudes.

\subsubsection{The origin of the S-swing of polarisation angle}

  \begin{figure}
   \includegraphics[width=0.48\textwidth]{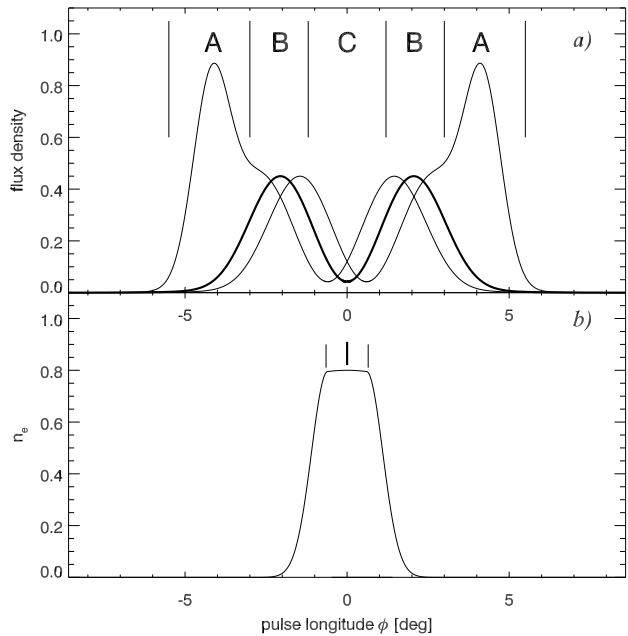}
       \caption{{\bf a)} Decomposition of a D-class profile into 
       elementary beams
       emitted from different locations in the plasma stream.
       The \emode-dominated beam (thick line) is emitted from the center,
and the \omode-contaminated, asymmetric beams from the outside edges
(thin lines). In the phase-interval
C the variations of polarisation angle are determined by the intrinsic
properties of the elementary beam (see text for more details).
{\bf b)} Plasma density distribution sampled by our line of sight.
The short bars mark the emission points for the curves in panel a.
       }
      \label{norvm}
   \end{figure}

The small width $\Delta\phi_m$ of the emitter in 
the magnetic azimuth 
implies that the projected direction of magnetic field does not change much
within the pulse window. Thus, our scenario implies almost no
change of polarisation angle (PA) due to the effect predicted by
the rotating-vector-model (RVM) of Radhakrishnan \& Cooke (1969).
The PA curve is predicted to stay flat only \emph{if} it is dominated 
by the RVM effect, ie.~if the observed PA is determined by the 
projection of $\vec B$
on the sky. If the PA is governed by non-RVM factors,
then behaviour that is \emph{in general different} from the RVM curve is
expected.

The non-RVM behaviour is indeed observed in several pulsars with double
peaks. For example, this is the case for B0834$+$26, B1919$+$21 (HR08),  
as well as for the geodetically-precessing pulsar B1913$+$16 (see
fig.~7 in Weisberg \& Taylor 2002).
However, the non-RVM origin \emph{seems} to be plain inconsistent 
with the paramount 
examples of the RVM exhibited by two well-known representatives of the D-class: 
B0301$+$19 and B0525$+$21. They are widely believed to present textbook cases
of the RVM-generated PA curve.

We claim that here again we become victims
of maliciously deceiving degeneration of Nature. 
The RVM model is really what
the name says: a curve determined by the sky-projection 
of a vector that rotates
under our line of sight. The projection of the circumpolar $B$-field lines
is just one example of such a vector. The other examples,
not even slightly less natural than the projected $\vec B$,
are the vectors of electric field in the radiation pattern surrounding
the instantaneous velocity of a relativistic particle.
We argue that the famous S-swing of PA in the B0525$+$21
(and other D pulsars) is mostly caused by intrinsic effects.
It is due to the cut of our sightline through 
the elementary (microscopic) \emode\ curvature beam.

A rather complicated code would be needed to prove it
numerically. Instead of that, we choose to delinate
the mechanism qualitatively. In Fig.~\ref{norvm}a
the observed pulse profile of a D-class pulsar is decomposed
into the elementary beams, emitted from three locations in the stream.
The thick curve presents the bifurcated
\emode-dominated beam emitted from the center
of the stream. The thin lines on the left- and right-hand side
present the beams emitted from the outside left and right edge of the beam,
respectively. The edge beams have their outside parts brightened by the
presence of the \omode.
Within the phase interval marked with the letter `A',
the polarisation properties are determined by relative
proportions of the two modes. If the segregation of the 
modes is not efficient in 100\%, a substantial elliptical
polarisation appears there (as is often observed). 
In the phase interval `B'
the flux observed at each phase 
always corresponds to the \emph{maximum} of different elementary beams,
emitted at different locations in the stream.
The changes of PA within the interval `B'
are therefore dominated by 
the changes of the $\vec B$ field direction.
Thus, it is in the phase interval B, where the emission from different
elementary beams is averaged out and can reveal the underlying geometry
of magnetic field. In this interval the observed average
PA curve can follow the RVM model. This can only happen if the
width of the phase interval B is noticeably large
(the width is determined by the relative
width of the stream and the opening angle $\psimax$ of the \emode\ beam).
The corresponding parts of the PA curve of B0525$+$21
are indeed very flat (and highly polarised because of the domination
of the \emode).

  \begin{figure}
   \includegraphics[width=0.48\textwidth]{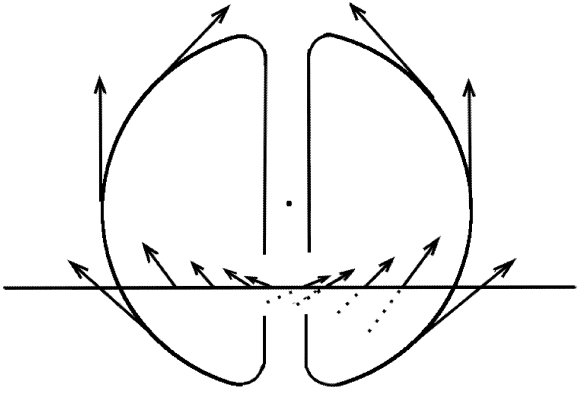}
       \caption{The origin of the PA swing in D-type radio pulsars.
A head-on view of the \emode\ curvature beam with the orientation
of polarisation vectors marked at selected places.
The passage of observer's line of sight
(horizontal line) 
leads to identical PA swing as in the case of the RVM model.   
   }
      \label{kleszcz}
   \end{figure}

In the phase interval `C' the situation changes completely:
for $\phi < 0$ the flux at each phase is now dominated by radiation from a 
single elementary beam. 
Our line of sight does not sample different beams emitted from different
locations in the stream. The sightline moves through the inner parts
of the elementary beam, and the changes of PA are determined
by the changes of polarisation direction intrinsic to the \emode\ beam
(note that in this phase interval the received radiation
is associated with a \emph{fixed} direction of $\vec B$).
For $\phi > 0$ (still in the C interval), 
the line of sight again samples 
a single elementary beam.
Therefore, in the innermost parts of the 
double conal profiles, the PA is determined by the intrinsic
polarisation 
properties of the \emode\ curvature beam. As shown in Fig.~\ref{kleszcz},
the orientation of the polarisation vectors (vectors of the electric field
$\vec E_w$ of the emitted wave) in the instantaneous \emode-beam
ensures the same variations of the PA as the projected magnetic field 
in the RVM model. This interpretation is based on the intrinsic
properties of the emitted radiation and assumes 
that they are not affected by any adiabatic walking below 
the polarisation-limiting radius (eg.~Cheng \& Ruderman 1979;
Melrose 1979; Lyubarsky 2002).

We thereby disclose a perverse situation, in which the central parts
of the observed PA curve, that resemble the RVM shape so closely,
have the origin that has nothing to do with the RVM idea.
On the contrary, the outer flat parts of the observed PA curve
(interval B) are more likely to reflect the underlying orientation of $\vec
B$ and to comply with the RVM origin.
Had this been the case, the difference of PA measured in the outer wings
of the profile $\Delta (\rm PA)$ would provide information about
 the azimuthal width of the stream: $\Delta\phi_m\sim\Delta(\rm PA)$
(provided the PA observed in the A and B region is not biased/dominated 
by intrinsic effects).

The same conclusion can be reached in a different way, by considering
the case of an infinitely-thin stream, and noting that the observed
PA curve should change continously for small increase of the stream width.
Specifically, for a stream of electrons flowing only along a single 
$B$-field line, the observed PA curve definitely has to be determined fully
by the intrinsic swing shown in Fig.~\ref{kleszcz}.
This is because the spatial extent of the emitter is a delta function
and there is no variation of the underlying direction of
$\vec B$. 
A small increase of the diameter of the stream must, by continuity,
produce only small changes of the initial PA curve. The increasing
spread of $\vec B$ within the stream of increasing thickness 
will therefore produce only a gradual change in the observed PA curve.
We therefore conclude that in general the observed PA curves of pulsars
are a convolution of intrinsic PA and $B$-field projection.

Since the opening angle of the \emode curvature beam
decreases with increasing frequency, the PA curve is expected
to depend on $\nuobs$, which is indeed the case for B0525$+$21
(and other objects).
Because the \emode-radiation has a minimum in the profile
center (Fig.~\ref{norvm}), any remnants of the \omode\ emission
can produce a minimum of polarisation degree there.
Even such detail is also observed in B0525$+$21 (eg.~fig.~1 in HR08).
In a more general case (than the primitive picture of Fig.~\ref{norvm}),
a large variety of PA curves is expected, depending on the 3D distribution
of plasma density in the stream, on the geometry of passage of
our sightline through the beam, on 
the ratio of both orthogonal modes,
and on the size of the elementary curvature beam. 
For realistic density distribution the phase intervals B and C can overlap
in phase, in which case the PA recorded at a given phase
can be partly determined by the intrinsic PA and partly by the projection of
$\vec B$. The subject deserves numerical study.

Interesting examples of intrinsic variations of PA curve
have also been observed under core components, where
the polarisation has been found to depend
on the subpulse brightness (B1237$+$25, Srostlik \& Rankin 2005;
B0329$+$54, Mitra, Rankin \& Gupta 2007). The core emission
has been found to be polarised orthogonally with respect to the
outer components (Srostlik \& Rankin 2005) which suggests
\omode\ origin of the core (Gil \& Snakowski 1990).

We conclude that the observed PA curve (at least for D-type pulsars)
is strongly influenced by the intrinsic
polarisation properties of the \emode-beam. Even in the most
RVM-like looking fragments it does not have the RVM orgin.
Though this interpretation invalidates the determination of the 
general geometric parameters ($\alpha$, $\zeta$) from the PA curves,
it opens a way to interpret their ubiquitous distortions from 
the RVM-shape.

\subsubsection{Profile width and the geodetic precession}

Precessing pulsars give us the unique opportunity to 
view their emission beams in two dimensions.
Especially precious are the pulsars in relativistic binary systems,
such as J0737$-$3039A/B, B1913$+$16, J1141$-$6545, B1534$+$12,
or J1906$+$0746.
They should exhibit the geodetic precession,
the rate of which is well determined by the general relativity.
For the traditional, axially-symmetric pulsar beams,
such precession should generally produce clear changes of profile width.
Several detailed studies of the phenomenon (Kramer 1998;
Clifton \& Weisberg 2008; Lorimer et al.~2006; Burgay et al.~2005)
have reported that the profiles refuse to change at the expected rate. 
This suspicious observation can be explained in several ways,
eg.~by near-alignment of the spin axis and orbital angular momentum,
or by the accidental approximate co-planarity of the rotation axis, 
orbital momentum,
and line of sight (see Clifton \& Weisberg 2008).
However, none of the explanations seems natural or likely,
especially in view of the increasing ubiquity of the problem.

  \begin{figure}
   \includegraphics[width=0.47\textwidth]{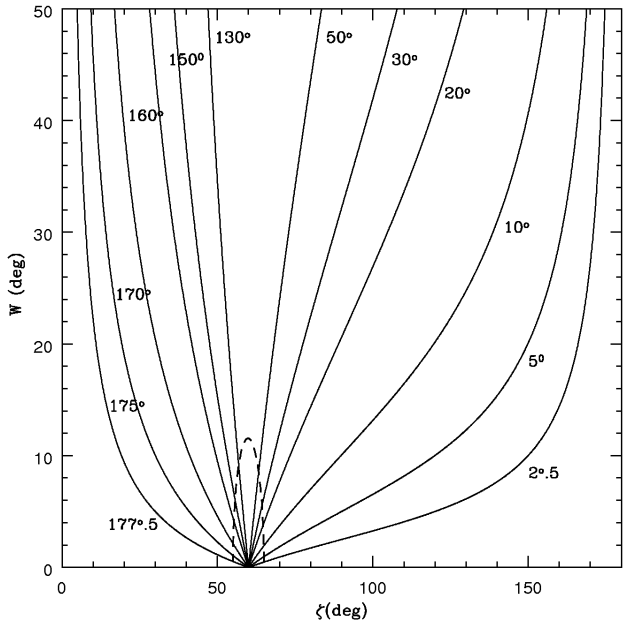}
       \caption{Variations of pulse width $W$ as a function
of the viewing angle $\zeta$ in the stream model 
(solid lines; numbers
give the value of $\phi_m=\Delta\phi_m/2$). The outermost two curves present
the thinnest streams of width $\Delta\phi_m = 5^\circ$ and $10^\circ$ in the
magnetic azimuth. The streams are centered at the main meridian 
and $\alpha=60^\circ$. Dashed line presents the standard case of a conal
beam with the opening angle $2\theta_m=10^\circ$.
   }
      \label{wbis}
   \end{figure}

We argue that the real reason for the profile stability is
the completely different topology of pulsar beams:
our sightline is cutting through the split/structured
\emph{fan} beams generated by elongated plasma
streams, not through the narrow cones.
The pulse profile of B1913$+$16 indeed has the D-class shape
and exhibits strongly distorted (non-RVM) PA curve
(fig.~7 in Weisberg \& Taylor 2002).
However, our fan-beam interpretation is likely to hold
not only for the D-class profiles. The pulsar J1906$+$0746
(Lorimer et al.~2006) belongs to a single-component class
(S type),
and exhibits interesting profile evolution:
despite its interpulse clearly appearing within a few-years-long
observation time, the width of the main pulse does not
vary considerably. A cut through fan-beams explains such
behaviour naturally.  

In the traditional conal model
the observed pulse width $W=2\phi$ is
modelled by assuming the fixed opening angle $\theta_m$ of the conal beam
in the equation:
\begin{equation}
\cos\phi = \frac{\cos\theta_m - \cos\alpha\cos\zeta}{\sin\alpha\sin\zeta}.
\label{cosphi}
\end{equation} 
In the fan beam scenario it is the
azimuth $\phi_m$ of a wedge-shaped stream 
that needs to be fixed, whereas $\theta_m$ in the above formula changes
with time, complying with the equation:
\begin{equation}
\cos\zeta =\cos(\pi-\phi_m)\sin\alpha\sin\theta_m + \cos\alpha\cos\theta_m.
\label{cozeta}
\end{equation}
The assumption of  $W=2\phi$ is only correct for a stream that 
is symmetric with respect to the main meridian (containing the rotation and
magnetic axes). In general, $\phi_m^{\rm le} \ne \phi_m^{\rm te}$
and the width is equal to $W=\phi(\phi_m^{\rm te}) - \phi(\phi_m^{\rm le})$,
with $\phi$ given by eq.~\mref{cosphi}.
The instantaneous value of $\cos\theta_m$ needs to be calculated
from the quadratic equation \mref{cozeta} and used in \mref{cosphi}.
This leads to the much slower variations of $W$ with $\zeta$, as can be seen
in Fig.~\ref{wbis} in which the solid lines show $W$ for streams
of various azimuthal width $\delta\phi_m = 2\phi_m$
and $\alpha = 60^\circ$.
The quasi-linear changes of $W$, characteristic for the observations of 
geodetically-precessing pulsars, are readily apparent for narrow streams
(outermost lines; see the values of $\phi_m$ marked at each line).
The changes of $W$ for a conal beam with $\theta_m = 10^\circ$
are shown for comparison with the dashed line.

Therefore, because of the geodetic precession
we have already traversed a long way
through the beam of B1913$+$16. The profile has not evolved much 
because we cut through a fan-shaped beam.
We argue that the stability of profile in both components of 
the double pulsar
J0737$-$3039A/B has the same origin, despite the profile classification
for the A star is far from clear.

\subsection{Sketch of a new classification scheme}

\begin{figure*}
   \resizebox{12cm}{!}{
\includegraphics[width=0.5\textwidth]{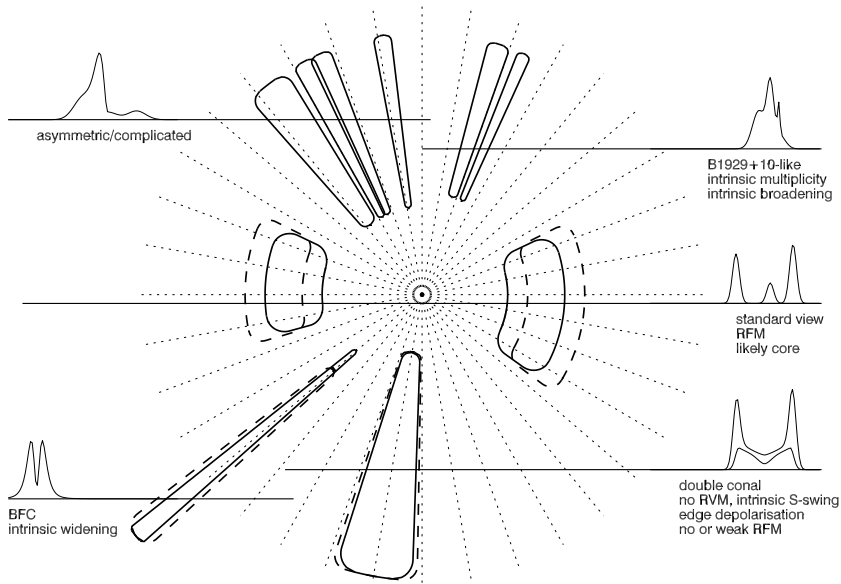}
}
      \hfill
      \parbox[b]{50mm}{
       \caption{A head-on view down the magnetic pole, presenting
the geometry of possible emission regions. Some of them are plotted at 
two frequencies: a lower one (dashed contours) and a higher one (solid
contours). The standard conal view of profile generation (middle sightline)
is contrasted with the stream origin that we advance in this paper
(bottom sightlines). The stream-cut origin
is likely not restricted to the D-class pulsars, as suggested in the upper
part of the figure.
       }
      \label{bcake}
      }
\end{figure*}

\subsubsection{Neither conal nor patchy: spoke-like, wedge-like, and
stream-like}

We have come to the conclusion that bifurcated components
in radio-pulsar profiles are produced by sightline cuts through
fan-beams of radiation from elongated streams.
There need to be stream-like emitters in the pulsar magnetosphere,
as shown in the bottom-left corner of Fig.~\ref{bcake}.
An important question is what is the length of the streams and 
what is the range of magnetic colatitudes that the fan beams subtend.
At least in some cases the fan-beams must extend quite
far from the dipole axis, to explain the large separation of bifurcated
components from the main parts of the profile (main pulse, interpulse).
The large extent could also explain the abundance of
pulsars with interpulses (eg.~Weltevrede \& Johnston 2008b).
If the streams are not exactly centered at the main meridian,
the main-pulse-interpulse separation can differ from $180^\circ$
as it sometimes happens to be observed (eg.~in B1055$-$52,
Weltevrede \& Wright 2009). 

The components of the BFC class are observed at frequency-independent
location in pulse profiles and widen with increasing $\nuobs$
roughly according to $\dbfc\propto\nuobs^{-1/3}$.

For the price of rejecting the RVM wisdom, we have also come to the
conclusion, that the double conal profiles (class D) are also
created by a cut through azimuthally-limited emitter, that is
a bit broader than in the BFC case
(right bottom corner of Fig.~\ref{bcake}).
The width of such profiles is mostly determined by the
fixed width of the stream. Any low-frequency widening
(of non-RFM nature) can result from density gradient
in the outskirts of the stream. The intrinsic $\nuobs^{-1/3}$
widening can also contribute to the apparent effect.

The fan-beam origin of pulse profiles is probably not limited
to the D class. The profile of B1929$+$10
is likely created by a cut through a structured stream, or through 
a few nearby streams. At 430 MHz, its profile has a pronounced 
emission component on the trailing side of the maximum, which
separates from the maximum at a rate roughly consistent 
with $\nuobs^{-1/3}$. Moreover, in the minimum between the component
and the central maximum, noticeable amount of orthogonal polarisation mode
appears (Rankin \& Rathnasree 1997).
This kind of corellation (appearance of another polarisation mode
at a minimum in total flux) could in principle be
produced by any process that removes the dominating mode.
However, here we find again that the \emode-dominated curvature beam
(case $\parallel$B in Fig.~\ref{etaperp}a) has the appropriate
intrinsic property to produce the phenomenon
(domination of the \omode\ at the minimum of the 
\emode).\footnote{Another beautiful example of such effect is seen in
B1745$-$12 (fig.~7 in Mitra, Sarala, \& Rankin 2004).}
We therefore suggest that the `multiple' appearance of the 
profile of B1929$+$10 is partially caused by the intrinsic
doubleness of the curvature beam, and the real number
of underlying streams is smaller than the number of observed 
components.
If the radiation in the pedestal emission
is averaged out to reveal the RVM polarisation angle, 
then this pedestal PA curve should be discordant
with the PA curve observed under the main pulse
(because the latter is dominated by intrinsic effects).
This likely explains the difficulty in fitting the
full-period interval with a single PA curve
(Rankin \& Rathnasree 1997; Everett \& Weisberg 2001).
B0950$+$08 may belong to the same, B1929$+$10-like class:
its profile tends to bifurcate at low $\nuobs$.
The extended leading-bridge of this profile is likely caused
by a stream that extends roughly tangentially
to the trajectory made by our line of sight. The bridge culminates
in an interpulse which again reveals doubleness 
(McLaughlin \& Rankin 2004).

A paramount example of stream-cut-generated profile
is provided by J0437$-$4715 in which the tendency to
`look double' is revealed both in the complicated
central parts and in the conal components.
Whereas the core can be produced in the usual
way (a cut through the dipole-axis-centered rings)
it may also be created by streams flowing near
the main meridian.
The outer ``conal" components
are definitely produced by streams with large $\phi_m$.
Deconvolution of this (and other) profiles
into the underlying emitting
structures (streams) can only be done if the bifurcated
nature of the elementary curvature beam is taken into account.

\section{Summary}

The bifurcated component of \jtt
can be comparably-well fitted with both the parallel-acceleration emission
and the curvature radiation. 
However, only the curvature radiation
can match the data simultaneously at two frequencies.
The two-planar topology of the \emode-dominated curvature beam
makes it very easy to 
reproduce the observed large depth of double notches.
It also explains the origin of bifurcated components
in terms of very natural geometry (sightline
cuts through plasma streams). 
This leads us to the firm conclusion that the observed coherent
radio emission from pulsars is the curvature radiation.

Components in radio pulsar profiles become wider, 
and finally bifurcate at low-frequencies because of the intrinsic
behaviour of the curvature beam. The scale of the bifurcation
makes it possible to determine the radius of curvature of electron
trajectory in the emission region, provided global geometry parameters
are known.

Double ``conal" profiles (D class) are not conal in the usual sense.
They are produced by the sightline cut through slightly broader
streams. 

The PA curves of such profiles are not determined by the projection
of the $\vec B$-field on the sky plane. Even when they resemble the
RVM shape (which they do rarely), 
the PA curves are mostly determined by the microphysical
polarisation structure of the \emode\ curvature pattern.
The S-swing created by a cut through such single pattern
is exactly the same as in the RVM case. Angular/spatial convolution of 
more patterns leads to a variety of observed PA curves.

The low-frequency broadening of D-type profiles
results from plasma density gradient at the outer surface of the stream,
and from the intrinsic broadening of the curvature beam.
It does not reflect the emission radius (it does not have the RFM origin). 

The narrow-stream geometry of the emitter implies fan-like shapes
of pulsar emission beams. This naturally
explains the extremely slow changes of pulse width due to the geodetic
precession.

The existence of fan-shaped beams increases the likelihood of observing
an interpulse. It also implies that
the number of pulsars in the Galaxy may be smaller
than estimated with the conal beams.

The results of this paper invalidate the traditional methods of 
determination of the radio emission
altitude at least for D-class pulsars.
This refers both to the geometrical method, and the one based on 
the relativistic-shift of PA curve.

The curvature-radiation beam with the partially-attenuated \omode\
has several features that are qualitatively 
consistent with many observed properties of
pulsars. The features include
the non-zero width of the curvature beam, the mirror symmetry with respect
to the B-field line plane, the `striped' (aziumuthally-separated)
polarisation structure, and the intrinsic 1:1 ratio of the orthogonal
polarisation modes at the beam edge.
They enable us to roughly comprehend the multiple imaging,
azimuthal separation of modes, edge depolarisation, and the polarisation
structure of D-type profiles.

The stream origin is likely 
for other types of pulsar profiles. In attempts to decipher
specific cases, it may be more efficient to think in terms
of spoke-like, or wedge-like emitter, not in terms of `conal or patchy'.

\section{Discussion}

Several coherency mechanisms for the amplification of the curvature
radiation had been proposed. They are usually divided into
the bunching processes, and the maser processes,
and both are reportedly problematic (Melrose 2006).

The elementary beam of the coherent curvature radiation, that we have
invoked from the fits of Section \ref{fits} and from the 
pulsar properties of Section \ref{appli} suggests that the coherency
is not due to the curvature-drift-driven maser (Zheleznyakov \& Shaposhnikov
1979; Luo \& Melrose
1992,
hereafter LM92)
nor due to the maser induced by the field-line torsion (Luo \& Melrose 1995).
It is true that the first model predicts the required two-planar
topology of the beam and the correct frequency dependence of $\dbfc$.
However, the maser implies intrinsically asymmetric beam
which does not seem to be consistent with the wonderful symmetry of the
double features. This can be seen
in fig.~1 of LM92. The amplified parts of the non-coherent
beam (determined by the negative absorption coefficient $\Gamma$
in the bottom-right panel) are located asymmetrically with respect
to the plane of electron trajectory (at the angles marked $\theta_1$
and $\theta_2$ in the figure of LM92). The maser operating thanks to the
$B$-field line torsion (Luo \& Melrose 1995) amplifies 
the radiation at extremely small angles with respect to $\vec B$. 
The bunching-induced coherency (eg.~due to the two-stream
instability, Ruderman \& Sutherland 1975) seems to have less problems
with the quasi-isotropic amplification of the non-coherent beam.
Recently this idea was studied by 
Gil, Lyubarsky \& Melikidze (2004)
in their version of `soliton bunching'. 
In their figure 3 they present
elementary beams of the \emode\ radiation that are fairly similar to the
BFC of \jttns. It is less clear if their mechanism can amplify comparable
amounts of the \omode\ in the outer parts of the beam.

The idea of ``fan" beams appeared from time to time in
pulsar literature. Eg.~Narayan \& Vivekanand (1983)
considered an elongated beam to model the notorious
profile of B0950$+$08. However, the beam elongation in their model
(Fig.~2 therein)
was still caused by a deformation (squeezing) of a conal beam
and did not have the stream-like topology.\\
It is peculiar that in the past the fan-beam origin of profile components
has not been emphasized sufficiently enough to spark 
development of a spoke-like model of pulsar beams. The intrinsic
fan-beam topology of the curvature radiation has been known from the early
days of synchrotron radiation theory, and
the tendency of curvature radiation to produce double components 
has even been incidentally illustrated 
in the literature (eg.~figs.~2 and 3 in Ahmadi \& Gangadhara 2002).

We conclude that the long-sought Rosetta Stone
needed to decipher the nature of pulsar radio emission
has finally been identified as the double features
in averaged pulse profiles.

\section*{acknowledgements}

JD is indebted to K.~Lazaridis, A.~Jessner,
R.~Manchester, and J.M.~Rankin
for providing us with pulsar data. He appreciates discussions with 
J.~Arons and M.~Kramer.
Comments of Geoffrey Wright,
our referee, considerably improved clarity of this paper.
This work was supported by the grant N203 017 31/2872
of the Polish Ministry of Science and Higher Education
and the Polish Astroparticle Network 621/E-78/SN-0068/2007.

\end{document}